\documentclass{elsart}

\usepackage{graphicx}
\usepackage{epsfig}

\usepackage{amssymb}

\begin{document}

\begin{frontmatter}

\title{Power laws in surface physics: \\
The deep, the shallow and the useful}

\author{Joachim Krug}

\address{Institut f\"ur Theoretische Physik, Universit\"at zu 
K\"oln \\ Z\"ulpicher Strasse 77, 50937 K\"oln, Germany}

\begin{abstract}
The growth and dynamics of solid surfaces displays a multitude
of power law relationships, which are often associated with geometric
self-similarity. In many cases the mechanisms behind
these power laws are comparatively trivial, and require little more
than dimensional analysis for their derivation. The information of 
interest to surface physicists then resides in the prefactors. This
point will be illustrated by recent experimental and theoretical work
on the growth-induced roughening of thin films and step fluctuations
on vicinal surfaces. The conventional distinction between trivial and nontrivial power
laws will be critically examined in general, and specifically in the context
of persistence of step fluctuations.  

\end{abstract}

\end{frontmatter}

\small 
\emph{Empirical science is apt to cloud the sight, and, by the very knowledge
of functions and processes, to bereave the student of the manly contemplation
of the whole. The savant becomes unpoetic.} \hfill Ralph Waldo Emerson

\normalsize

Per Bak's theory of self-organized criticality is based on the
observation that power laws are ubiquitous in nature, and the understanding
that this observation
\emph{requires a (general) scientific explanation}.
The latter point is often taken for granted in statistical physics,
but it is less self-evident in other disciplines. In fact, it could be
argued that the predelection of statistical physicists for power
law relationships is a professional deformation, which originates
in the success story of the theory of equilibrium critical phenomena.
In that context, power laws are indeed anomalies which require the
ingenious machinery of the renormalization group for their explanation.

However, not every power law carries a deep message. Statistical physicists
account for this by distinguishing between \emph{trivial} and
\emph{nontrivial} power laws. Though deeply rooted in our jargon, this
distinction hard to make precise. Most people would agree that the
relation 
\begin{equation}
\label{RW}
\langle \Delta x^2 \rangle \sim t
\end{equation}
for the mean square
displacement of a random walker (really a manifestation of the central
limit theorem) is a trivial power law, whereas, say, the Onsager exponents
for the two-dimensional Ising model are nontrivial. But consider
Kolmogorov's 1941 theory of fully developed turbulence \cite{Frisch}.
The derivation of the $k^{-5/3}$-energy spectrum is trivial, in the
sense that it uses only dimensional analysis, but it requires
the higly nontrivial physical insight that energy dissipation is
constant across scales. Clearly the demarcation line between the 
trivial and the nontrivial also evolves with the progress of science.

Since the discovery of the fractal nature
of diffusion-limited aggregation (DLA) more than two decades ago 
\cite{Witten81}, scaling concepts have become one of the
main tools in the study of growth processes in condensed
matter \cite{Barabasi95,Krug97b,Meakin98,Michely03}. While much of the
theoretical activity has been driven by the 
(still unfinished) quest of understanding the two most prominent
nontrivial power laws in the field --  the fractal dimension of
DLA, and the strong coupling exponents of the Kardar-Parisi-Zhang (KPZ)
equation \cite{Kardar86} in dimensions larger than one 
\cite{Krug97b,Krug91,Halpin95} -- experimentalists have begun
to routinely employ the concepts of scaling and self-similarity
to analyse topographic data in thin film and crystal growth.    
In the following I will describe some recent applications of scaling
ideas in surface physics. I will argue that, more often than not,
power laws that, by the standards of statistical physicists,
are quite trivial, have been of most use in interpreting experimental 
data and gaining insight into kinetic processes at real surfaces.
The discussion will include 
the currently popular concept of persistence of a stochastic process 
\cite{Majumdar}, which provides an interesting perspective on the
distinction between trivial and nontrivial power laws.

\begin{figure}
\centerline{\includegraphics[width=0.8\textwidth]{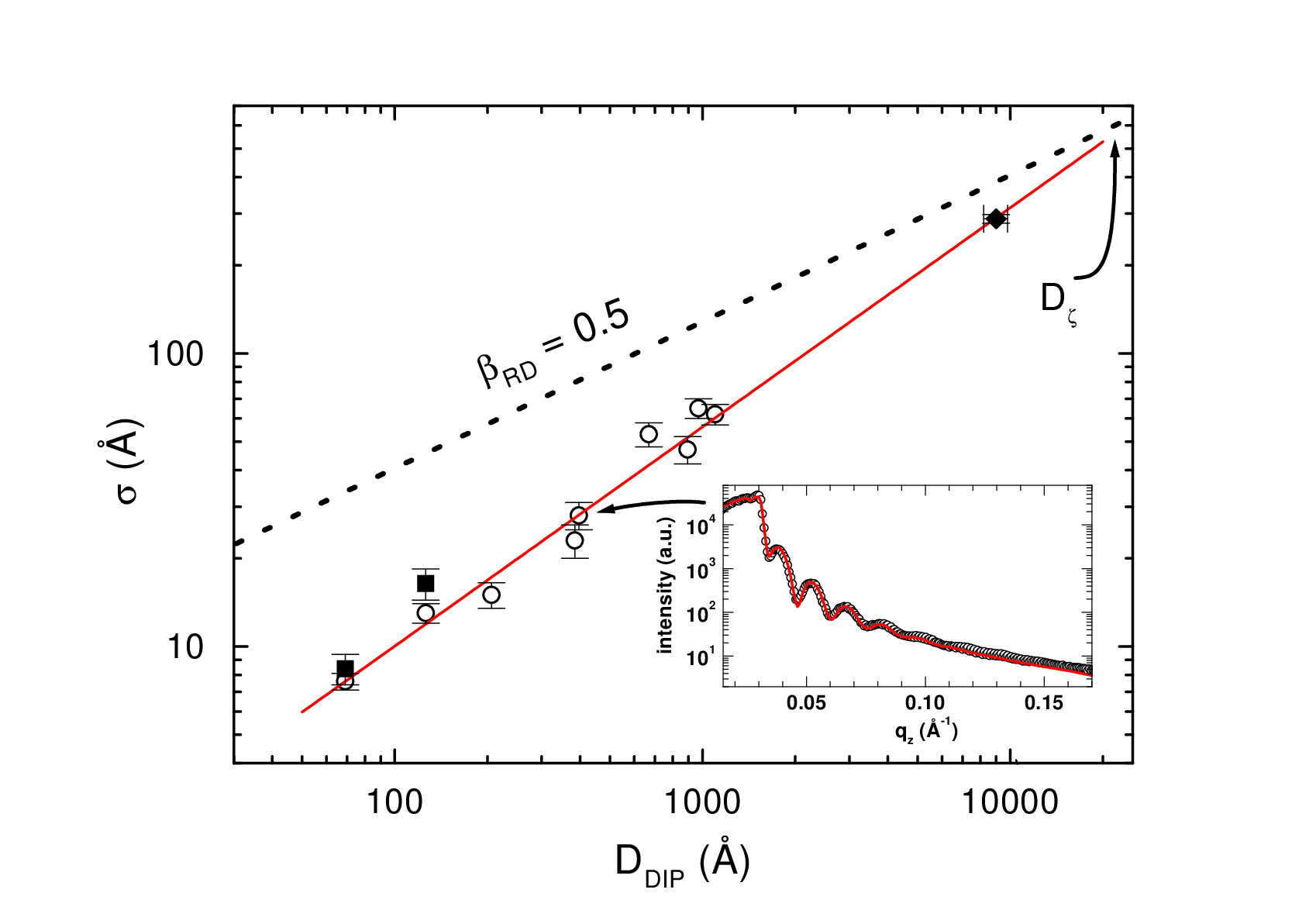}}
\caption{Experimental data for the surface roughenss $\sigma$
as a function of DIP film thickness. Full data points
were obtained from atomic force microscopy, and open data
points from x-ray diffraction. Inset shows a typical
fit of the measured scattering intensity to the form
expected for a rough surface (from \cite{Duerr03})}
\label{Duerr}
\end{figure}

Our first example is the growth-induced roughening of thin films
of an organic semiconductor, diindenoperylene (DIP) \cite{Duerr03}.
The relevant scaling law is the relationship between the 
root mean square surface roughness $\sigma$ (the standard deviation 
of the film height distribution) and the film thickness $D$,
\begin{equation}
\label{sigma}
\sigma(D) \sim D^\beta,
\end{equation}
which defines the roughening exponent $\beta$. The experimental data for
$\sigma(D)$ shown in Fig.\ref{Duerr} are remarkable in several respects.
First, they cover more than two decades in film thickness, and they
contain consistent results obtained using two completely different methods,
x-ray diffraction and atomic force microscopy. The second remarkable
feature is the experimentally determined value $\beta = 0.75 \pm 0.05$
of the roughening exponent. This value \emph{exceeds} the random deposition
(RD) limit $\beta_{\mathrm{RD}} = 1/2$, which applies in the absence of
any mass transport between different layers of the growing film 
\cite{Barabasi95,Michely03}. In this limit the height follows 
a Poisson distribution, and (\ref{sigma}) becomes equivalent to 
the random walk relation (\ref{RW}). Random deposition can be approximately
realized in the growth of crystalline metal films,
where transport between atomic layers is inhibited by step 
edge barriers (see below).   
Since mass transport processes along the growing surface are driven
by differences in bonding energy, they will generally tend to smoothen
the film. The random deposition roughness 
$\sigma_{\mathrm{RD}}$ would therefore be expected
to define an upper bound on the roughness that random fluctuations
can induce during the growth of a thin film; in particular, 
$\beta \leq \beta_{\mathrm{RD}} = 1/2$. In this sense systems
with $\beta > 1/2$ are anomalous, and constitute examples 
for the (largely unexplained) phenomenon of \emph{rapid roughening}
\cite{Krug97b,Duerr03}.  

But do the DIP data in 
Fig.\ref{Duerr} really violate the RD bound? To answer this question,
it is essential to include also the prefactor of the power law
relationship (\ref{sigma}) in the analysis. Denoting by $d$ the
thickness of a single atomic layer, the random deposition 
roughness is $\sigma_{\mathrm{RD}} = d \sqrt{D/d}$, which is depicted
as a dotted line in Fig.\ref{Duerr}. The comparison shows that the
experimental roughness data remain \emph{below} the random deposition
limit even for the largest film thicknesses. It is conceivable
(though perhaps not very likely) that the roughening exponent crosses
over to a value below 1/2 before the critical thickness $D_\zeta$
in Fig.\ref{Duerr} is reached, and that the early time value 
$\beta \approx 0.75$ is merely a transient\footnote{Similar transients
have been observed in simulations of metal epitaxy \cite{Caspersen02}.}; 
unfortunately films with thicknesses $D > D_\zeta$ could not be grown for
technical reasons. If, on the other hand,
the measured value $\beta \approx 0.75$ is truly asymptotic, then
the observed roughness cannot be due to the random fluctuations in the
deposition beam. In \cite{Duerr03} it was conjectured that quenched
in-plane disorder arising from the boundaries of the tilt domains
in the organic thin film could be responsible for the rapid roughening
behavior. Such disorder is known to induce sublinear roughening behavior
of the form $\sigma \sim D/\ln(D)^\psi$, which can mimic a power law with
exponent $1/2 < \beta < 1$ over extended time scales \cite{Krug95}. 
Unfortunately, the available 
experimental information about the growth process
on the molecular level is insufficient at present to validate or refute
this hypothesis. 
  
If our understanding of DIP film growth, as described above,
seems unsatisfactory, it is nevertheless fairly representative of the field as a whole.
Although the power law relationship (\ref{sigma}) has been, and is currently
being reported for a host of growth systems \cite{Barabasi95,Meakin98}, including metallic as well
as semiconductor materials and crystalline as well as amorphous films, and 
although the measured values of $\beta$ tend to cluster around numbers
that can be derived from theoretical models \cite{Krim95}, a clear-cut example, where
a well-understood growth mechanism gives rise to a definite prediction for
$\beta$ which is quantitatively confirmed by experiment, is so far missing.
In particular, 2+1--dimensional KPZ scaling, a mathematical object under passionate
theoretical pursuit for close to twenty years, is still to emerge in any real growth
system\footnote{In 1+1 dimensions KPZ-scaling with the (trivial or nontrivial?) exponent $\beta = 1/3$
has been demonstrated for one-dimensional slow combustion fronts in paper \cite{Maunuksela97}.}.

Perhaps the only exception to this statement
is mound formation in homoepitaxial crystal growth \cite{Michely03},
where the exponent $\beta$ describes the roughening of a morphology that, rather than
being scale-invariant, displays a distinct lateral length scale.
The growth of ``wedding cakes'' under conditions of 
strongly inhibited interlayer transport is particularly simple
\cite{Krug97,Kalff99}. In this limit the roughening exponent takes on the random deposition
value $\beta = \beta_{\mathrm{RD}} = 1/2$, while deviations of the height distribution
from the ideal Poisson form show up in the prefactor. 
The Poisson distribution is cut off at large heights because a new layer can be nucleated
on top of a mound only when the top terrace has reached a critical size
\cite{Krug00}. This leads 
to the expression       
\begin{equation}
\label{wed}
\sigma/d = \sqrt{(1 - \theta_c) D/d} 
\end{equation}
for the surface roughness, where $\theta_c$ is the coverage corresponding to 
the critical top terrace size \cite{Krug02}. Equation (\ref{wed}) is our first example
of a trivial power law where the nontrivial information (the specifics of the 
interlayer transport processes that determine $\theta_c$ \cite{Krug00,Krug02}) resides
in the prefactor.  

In the remainder of the paper we will be concerned with the roughening
of \emph{one-dimensional} objects (lines). Rough lines appear naturally
as atomic steps on
vicinal crystal surfaces \cite{Jeong99,Giesen01}. A vicinal surface is obtained
by cutting the crystal in a direction close to (in the \emph{vicinity} of)
a high symmetry plane, and it consists of high symmetry terraces separated
by steps of monoatomic height. 
In thermal equilibrium, the steps are roughened by thermal fluctuations.
We describe the step at time $t$ by a function $y(x,t)$,
where the $x$-axis is taken along the mean step direction and the $y$-direction
is perpendicular to the step (in the direction of vicinality). Up to the 
length scale where collisions with neighboring steps become important
the static step conformation is the graph of a one-dimensional random walk,
$\langle (y(x,t) - y(x',t) )^2 \rangle \sim \vert x- x' \vert$.  
Time-dependent fluctuations are well described by a
linear Langevin equation of the form 
\begin{equation}
\label{Langevin}
\frac{\partial}{\partial t} y(x,t) = - K \left( - \frac{\partial^2}{\partial x^2} 
\right)^{z/2} y(x,t) + \eta(x,t)
\end{equation}
where $K$ is a positive constant, $\eta(x,t)$ is white noise, and the 
\emph{dynamic exponent} $z$ depends on the dominant kinetic pathway through
which the step fluctuations relax to equilibrium. The most important
cases are $z=2$, corresponding to fast mass exchange between the step and
the terrace (\emph{nonconserved} kinetics) and $z=4$ corresponding to 
mass transport only along the step (\emph{conserved} kinetics). It is straightforward
to show from (\ref{Langevin}) that the temporal step correlations scale as
\begin{equation}
\label{C}
C(t) = \langle (y(x,s) - y(x,s+t))^2 \rangle \sim t^{1/z}.
\end{equation} 
By analogy with (\ref{sigma}) we can say that 
the roughening exponent of the step is $\beta = 1/2z$.

\begin{figure}
\includegraphics[width=0.5\textwidth]{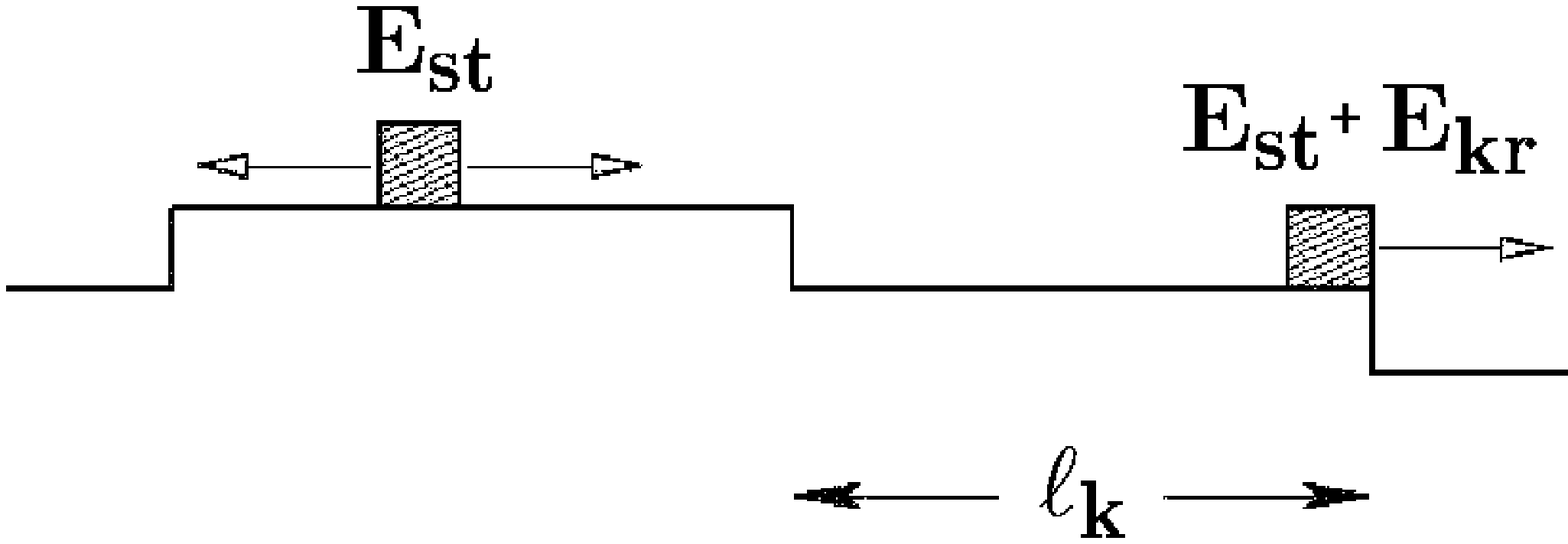}\hspace*{1.cm}\includegraphics[width=0.35\textwidth]{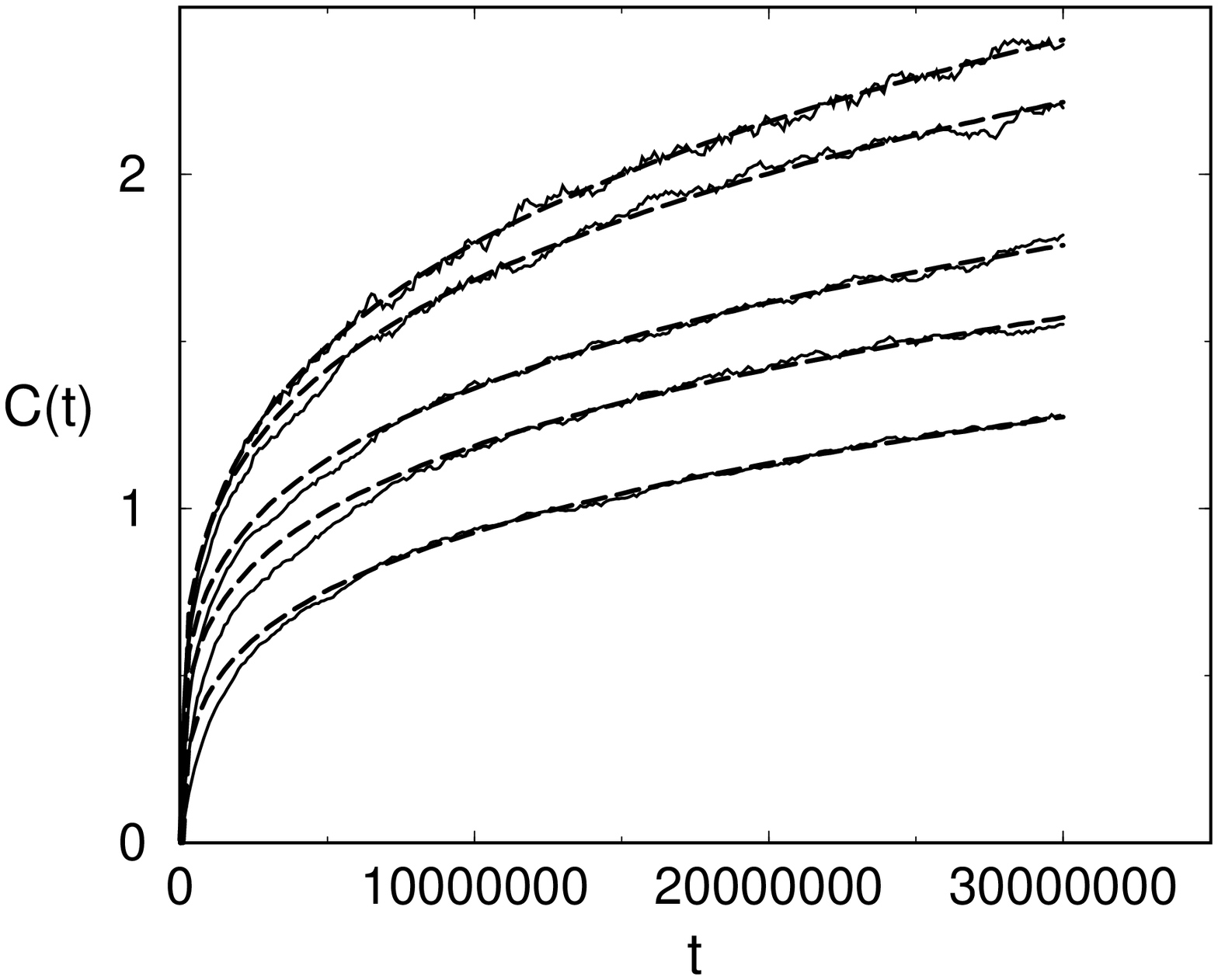}
\caption{Left panel: Schematic of atomic processes at a step edge. The 
diffusion of single atoms along straight segments of the step is governed
by an energy barrier $E_{\mathrm{st}}$. The detachment from a kink requires
the additional energy $E_{\mathrm{det}}$, and the rounding of the kink 
the energy $E_{\mathrm{kr}}$. The distance between kinks 
is $\ell_{\mathrm{k}}$. Right panel: Temporal step correlation function
$C(t)$ obtained from Monte Carlo simulations for different values of 
$E_{\mathrm{kr}}$. Dashed lines are fits of the form $C(t) = A t^{1/4}$
(from \cite{Kallunki03})}  
\label{Jouni}
\end{figure}

The scaling law (\ref{C}) has been observed in many experiments, on both metal
and semiconductor surfaces \cite{Jeong99,Giesen01}. The focus in these experiments
has usually not been on the power laws as such, which (in the sense of 
our introductory discussion) are rather trivial manifestations of the simple
linear dynamics (\ref{Langevin}). The relation
(\ref{C}) is useful mainly as a classification
tool, which serves to identify the dominant step relaxation processes.
The nontrivial (and materials-specific) information lies instead in the
temperature dependence of the prefactor, which gives insight into the
energy barriers governing the atomic processes at the step edge.   
Examples of such processes are shown in the left panel of Fig.\ref{Jouni}.
The right panel shows simulation data for the correlation function 
$C(t)$, which illustrate the decrease of the prefactor of the $t^{1/4}$-law
as the kink rounding barrier $E_{\mathrm{kr}}$ is increased. As shown by the
detailed analysis in \cite{Kallunki03}, the activation energy of the prefactor
depends linearly on $E_{\mathrm{kr}}$, provided this quantity is larger
than the kink energy (the energy cost for the formation of a kink), and
therefore a temperature-dependent measurement of $C(t)$ can be used to 
experimentally determine $E_{\mathrm{kr}}$.

It would be premature to conclude, however, that the kinetics of step fluctuations
can be reduced completely to ``trivial'' power laws like (\ref{C}). Although the
Langevin equation (\ref{Langevin}) is linear, it is still considerably more
complex than a simple random walk, because the step is a spatially extended
object. As a consequence, the step position $y(x,t)$ at fixed $x$ is a 
\emph{non-Markovian} process in $t$. The non-Markovian character of a stochastic
process manifests itself clearly when considering the \emph{persistence
probability}, i.e. the probability that the process does not cross a
particular value     
in a specified time interval. 
The computation of this quantity requires the knowledge of temporal correlations
of arbitrary order. For a Gaussian process, such as the solution of the
Langevin equation (\ref{Langevin}), all higher order correlation functions are,
in principle, encoded in the two-point function (\ref{C}), but in practice
the calculation of the persistence probability for a general Gaussian process
is a hard, unsolved problem \cite{Majumdar}.

\begin{figure}\includegraphics[width=0.5\textwidth]{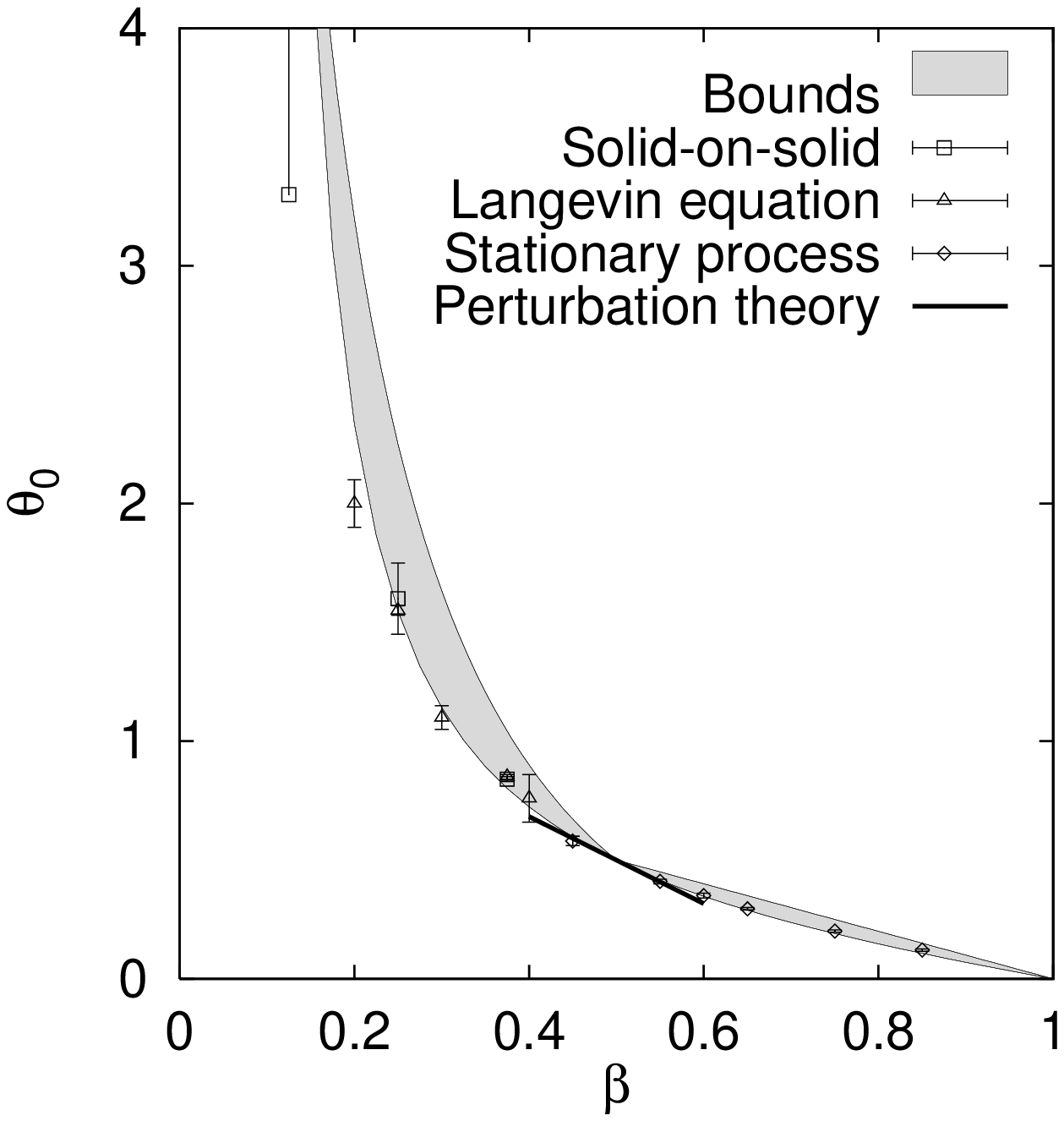}
\includegraphics[width=0.5\textwidth]{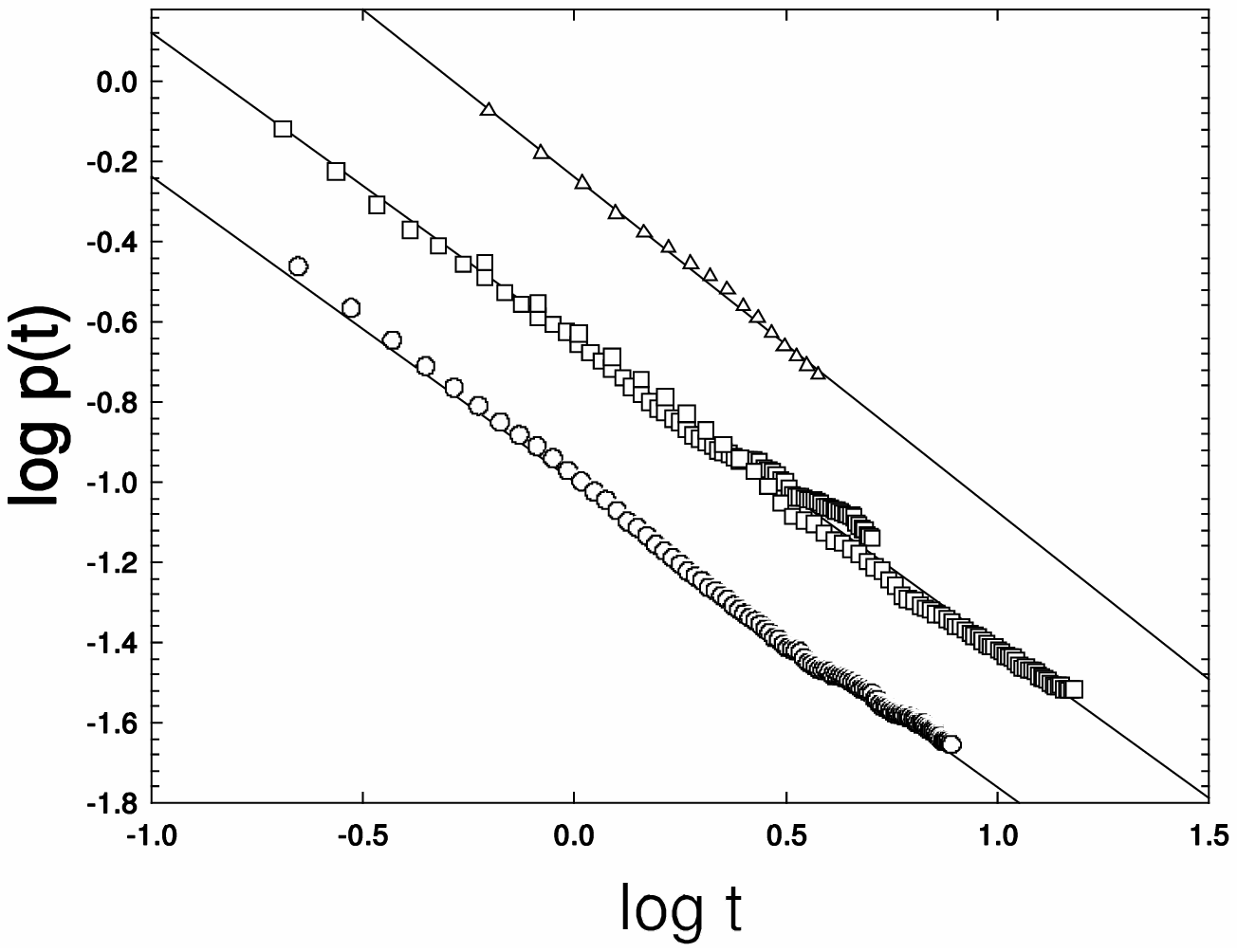}
\caption{Left panel: Transient persistence exponent $\theta_0$ as a function
of the roughening exponent $\beta$ for the linear Langevin equation (\ref{Langevin}).
The figure shows numerical estimates obtained from 
discrete solid-on-solid models (squares),
direct simulation of the Langevin equation (triangles), simulation of an
equivalent stationary Gaussian process (diamonds), a perturbation expansion
around the Markovian case $\beta = 1/2$ (bold line) and rigorous bounds
derived by comparison with related Markovian processes (gray area)
(from \cite{Krug97c}). Right panel: Experimentally determined persistence
probability for steps on a Si(111)-surface which has undergone a 
reconstruction by deposition of a fraction of a monolayer of Al. Symbols
show measurements at 970 K (squares), 870 K (circles) and 770 K (triangles).
The data sets have been shifted vertically for greater clarity; in fact   
the data show no systematic temperature dependence 
(from \cite{Dougherty02})}
\label{Pers}
\end{figure}

To be concrete, we define
the step persistence probability 
\begin{equation}
\label{Persistence}
P(t_0, t) = \mathrm{Prob}[y(s) \neq y(t_0) \vert t_0 \leq s \leq t_0 + t],
\end{equation}
where it is understood that the step is completly straight at time 
$t=0$ ($y(x,0) \equiv 0$). Note that here the specified level that the step should not
cross is set by its position at the beginning of the time interval. 
The analytic and numerical investigations in \cite{Krug97c} have shown
that the quantity (\ref{Persistence})
displays two distinct scaling regimes. In the \emph{transient}
regime $t_0 \ll t$ the persistence probability decays as $P \sim t^{-\theta_0}$,
while in the \emph{steady state} regime $t_0 \gg t$ it decays as
$P \sim t^{-\theta_S}$. The two persistence exponents $\theta_0$ and 
$\theta_S$ are different. Moreover, while the steady state exponent is simply
related to the roughening exponent $\beta$ by 
\begin{equation}
\label{thetas}
\theta_S = 1 - \beta,
\end{equation} 
the 
transient exponent is truly ``nontrivial'', in the sense that it can be computed
only approximately using a range of fairly sophisticated methods 
(see Fig.\ref{Pers}). 

The relationship (\ref{thetas}) provides a good
illustration of the fallibility of the distinction between trivial and 
nontrivial scaling exponents. On the one hand it would seem that $\theta_S$, being
simply related to the arguably trivial exponent $\beta$, should also be
classified as trivial. On the other hand the relation (\ref{thetas}), while
well supported by simulations and plausible scaling arguments,
is not rigorously established; in fact, in the context of the general
problem of determining the persistence probability of a Gaussian process
with known correlator, there is no reason why 
$\theta_S$ should be simpler to compute, and hence less nontrivial,
than $\theta_0$. The scaling relation (\ref{thetas}) holds because the 
stochastic process $y(x,t)$ at fixed $x$, in the steady state regime, combines
two invariance properties: It is scale invariant \emph{as well as} translationally
invariant in time\footnote{Together these two properties define a
Gaussian process known as \emph{fractional Brownian motion} (fBm) \cite{Mandelbrot68}, 
a generalization of the Wiener process, the continuum limit of the simple random walk.
For another recent application of fBm to a persistence problem
see \cite{Majumdar03}.}\cite{Krug98}. 
Because it is based on these general principles,
(\ref{thetas}) appears to hold also for non-Gaussian interface fluctuations, such as
those of the KPZ universality class, where not even the temporal two-point correlations
(let alone correlations of higher order) are explicitly known \cite{Kallabis99}.   

Several groups have recently met the challenge of experimentally measuring the 
persistence probability of one-dimensional interfaces, for steps on vicinal surfaces 
\cite{Dougherty02,Dougherty03,Constantin03} as well as for slow
combustion fronts in paper \cite{Merikoski03}. In all cases the steady state persistence
exponent $\theta_S$ was measured, and the scaling relation (\ref{thetas}) was confirmed,
with $\beta = 1/4$ \cite{Dougherty02}, $\beta = 1/8$ \cite{Dougherty03} and 
$\beta = 1/3$ \cite{Merikoski03}, respectively.
The transient persistence exponent (which is of greater theoretical interest) has so far
not been experimentally accessible, because of the difficulty of preparing the special
(flat) initial condition that it refers to. In addition to the persistence probability
defined by (\ref{Persistence}), in \cite{Dougherty02} also the probability for the
step not to cross the \emph{mean} step position (determined as an average over the
measured time series) was considered. This quantity turns out \emph{not} to follow
a power law. This is because in the thermodynamic limit the step position starts,
with probability one, infinitely far away from the mean (the variance of $y$ diverges
with diverging step length); hence the persistence probability remains at unity
and decays exponentially only on a much larger time scale set by the finite length
of the step \cite{Dasgupta03,Satya04}. 

Experimental step persistence
data from \cite{Dougherty02} are displayed
in the right panel of Fig.\ref{Pers}. In contrast to the step correlation function in Fig.
\ref{Jouni}, the persistence data show no systematic temperature dependence; the
prefactor of the observed power law depends only on the sampling time \cite{Williams}.
This is because the property of a height fluctuation to return (or not) to its
initial value in a prescribed time interval does not depend on the overall 
amplitude of the fluctuation; the persistence probability is a functional
only of the \emph{shape} of the (suitably normalized)
correlation function \cite{Krug97}. As a consequence, persistence measurements
cannot be used to extract energy barriers for specific microscopic processes.
The benefit of these measurements is of a more fundamental nature --
they prove that the theoretical description of step fluctuations through the Langevin
equation (\ref{Langevin}), and the underlying picture
of ``universality classes'' encoded by the dynamic exponent $z$, 
extends beyond the two-point function to correlations of arbitrary order.   
As the persistence probability is known to be extremely sensitive to hidden
temporal correlations affecting the interface fluctuations \cite{Kallabis99},
it may be particularly useful in cases where the assignment of the universality
class (the value of $z$) through more conventional measurements is ambiguous.

Where does this leave us with regard to the general reflections on power laws, the
trivial and the nontrivial, the deep and the shallow, which introduced this paper?
From discussions in the early 1990's, when SOC was challenged by the competing
concept of Generic Scale Invariance 
\cite{Grinstein90}, I recall that Per Bak had a distinct,
and rather unfavorable opinion of simple power laws generated by simple equations
such as (\ref{Langevin}): He referred to them as systems operating by the 
\emph{garbage in, garbage out} principle, because they merely perform a (linear, or,
more generally, nonlinear) transformation of the driving white noise into correlated
fluctuations. Per viewed self-organization as an essential part of SOC. He emphasized
the capability of self-organizing systems to develop new, emergent levels of 
structure, and to undergo a history which is open to contingent influences.
Is a rough surface or a fractal cluster grown by DLA self-organized in this sense? 
Probably not\footnote{Nevertheless it is true that scale-invariant structures
need sufficient time to develop; more precisely, the time $T$ required to grow a structure
of size $R$ typically scales as $T \sim R^z$ with $z > 1$. This is probably
the reason why most spatial power laws in Nature extend over less than two
orders of magnitude \cite{Avnir98}: Under typical physical conditions, there is
insufficient time for scale-invariant correlations to develop further.}. 
Still, we must accept that many of the power laws in the world
that surrounds us have rather humble origins. As theoretical physicists, our task
it to explain these origins; but as ``experimental philosophers'' (Per Bak) we would
also like to know what they \emph{mean}. The latter question 
is of course not one to be answered, but one that is to be constantly 
clarified (and reobscured) in the ongoing discourse of the community. 
In these discussions Per will be sorely missed, and 
gratefully remembered for a long time to come.

\section*{Acknowledgements}
I wish to thank Satya Majumdar for helpful correspondence,
and Arndt D\"urr and Ellen Williams
for the permission to reproduce experimental data.
The work reported here was conducted at the University of Duisburg-Essen
with partial support of DFG within SFB 237.

\end{document}